\documentclass[a4paper,twocolumn,pra,showpacs,showkeywords,aps,nofootinbib]{revtex4}
\newcommand{\beq}{\begin{equation}}
\newcommand{\eeq}{\end{equation}}
\newcommand{\beqa}{\begin{eqnarray}}
\newcommand{\eeqa}{\end{eqnarray}}

\def\ra{\rangle}
\def\la{\langle}

\usepackage{amsmath,amsfonts,amssymb}
\usepackage{graphicx}
\DeclareGraphicsExtensions{.pdf,.png,.jpg}
\usepackage{epstopdf}
\usepackage{bm}
\usepackage{dsfont}

\usepackage{color}

\begin{document}
\title{Fast transport of two ions in  an anharmonic trap}

\author{M. Palmero$^{1}$}
\author{E. Torrontegui$^{1}$}
\author{D. Gu\'ery-Odelin$^{2}$}
\author{J. G. Muga$^{1, 3}$}

\affiliation{$^{1}$Departamento de Qu\'{\i}mica F\'{\i}sica, Universidad del Pa\'{\i}s Vasco - Euskal Herriko Unibertsitatea, 
Apdo. 644, Bilbao, Spain}
\affiliation{$^{2}$Laboratoire de Collisions Agr\'egats R\'eactivit\'e,
CNRS UMR 5589, IRSAMC, Universit\'e de Toulouse (UPS), 118 Route de
Narbonne, 31062 Toulouse CEDEX 4, France}
\affiliation{$^{3}$Department of Physics, Shanghai University, 200444 Shanghai, People's Republic of China}
\begin{abstract}
We design  fast trajectories of a trap to transport two ions  using  a shortcut-to-adiabaticity technique based on invariants. 
The effects of anharmonicity are analyzed first  perturbatively, with an approximate, single relative-motion mode,  description.
Then we use classical calculations and 
full quantum calculations.  This allows to identify discrete transport times that  minimize excitation
in the presence of anharmonicity. An even better strategy to suppress the effects of anharmonicity 
in a continuous range of transport times is to modify the trajectory using an effective trap frequency
shifted with respect to the actual frequency 
by the coupling between relative and  center of mass motions.      
\end{abstract}
\maketitle
%
%
\section{Introduction}
Quantum information processing based on trapped ions may 
be applied to a large number of qubits
(and become scalable)  
by moving the 
ions between fixed zones where logic operations are performed \cite{q.computer,Rowe,Wineland,simulator}.  
The transport should be fast but excitations should also be avoided at the destination site. 
Different approaches have been proposed to implement faster-than-adiabatic  transport of cold 
atoms \cite{Uli,David,Calarco,fast-forward,Erik,Xi,BECtransport}. Diabatic transport of cold neutral atoms was demonstrated by D. Gu\'ery-Odelin and coworkers \cite{David} and, recently, 
fast transport of single or two trapped ions was also realized by two groups 
\cite{Bowler,Schmidt-Kaler,Roos}. 
One of the proposed approaches makes use of invariants to 
design trap trajectories without final excitation \cite{Erik,Xi,BECtransport,review}. It is very flexible and provides  by construction,
under specific conditions, a motionally unexcited final transported state. It also allows  for 
further trajectory optimization taking into account different experimental constraints, and robustness versus noise 
\cite{Andreas}. 
The invariant-based inverse engineering method has been applied so far to model 
the fast transport of a single particle \cite{Erik,Xi} and Bose-Einstein condensates \cite{BECtransport}.
In this paper, we extend the theoretical analysis in \cite{Erik} to two Coulomb-interacting particles within a single trap, focusing
on the effects of a mild anharmonicity which is present in any experimental setting \cite{Uli,anh}. 
In Section \ref{sec2} we study the transport of two ions first in a harmonic trap and then in  an anharmonic trap with an added time-dependent linear 
potential to compensate the inertial force.  
The applicability of this compensating method may be limited so other options  are explored. In particular we consider in Section \ref{sec3} 
the effect of anharmonicity when the trap trajectories are designed for an unperturbed (harmonic) trap.  This is done using an approximate 
one dimensional (1D) theory combined with perturbation theory. In Sec. \ref{sec4} we study numerically the full two-dimensional (2D) problem.  
The article ends with a discussion in Sec. \ref{sec6} and an Appendix on the extension of some of the  results to the transport of  $N$ ions. 
%
%
%
%
%
\section{Two-ion transport\label{sec2}}
\subsection{Harmonic Trap}
Let us examine first the transport of two single-charge ions of mass $m$ in an effectively one dimensional  harmonic trap that moves from $0$ to $d$
in a time $t_f$.  
Let $\sf{q}_1$ and $\sf{q}_2$ be the coordinates of the two ions with momenta $\sf{p}_1$ and $\sf{p}_2$ 
and $Q_0(t)$ the trajectory of the trap minimum. We use first the laboratory frame and make by now no formal distinction between 
operators and $c$-numbers, since the context will make clear the meaning of the symbols (the exception is Sec. IIC). 
The Hamiltonian includes a kinetic term, a harmonic potential, and an interaction potential due to the Coulomb force,
\beqa
H&=&\frac{\sf{p}_1^2}{2m}+\frac{\sf{p}_2^2}{2m}+\frac{1}{2}m\omega^2[({\sf{q}_1}-Q_0)^2+({\sf{q}_2}-Q_0)^2]
\nonumber\\
&+&\frac{C_c}{\sf{q}_1-\sf{q}_2}.
\label{H2har}
\eeqa
$\omega/(2\pi)$ is the trap frequency and  $C_c=\frac{e^2}{4\pi\epsilon_0}$, where $e$ is the electron charge and $\epsilon_0$ 
the vacuum  permittivity.  Here and below we omit frequently the time argument of the trap position, i.e., $Q_0=Q_0(t)$. 
We set $\sf{q_1}>\sf{q_2}$ because of the strong Coulomb repulsion. 
The  wavefunctions of the ions never superpose, so  we may 
effectively  treat the particles as distinguishable and the symmetrization of the  wavefunction
would not provide any new physical effect. This assumption is largely accepted 
when interpreting current experiments.  

Let us now introduce coordinates and momenta for center of mass (CM) and  relative motion,
\beqa
\begin{aligned}
Q&=&\frac{1}{2}(\sf{q_1}+\sf{q_2});\, P=\sf{p_1}+Ê\sf{p_2},
\\
r&=&\frac{1}{2}(\sf{q_1}-\sf{q_2});\, p=\sf{p_1}-\sf{p_2}. 
\end{aligned}
\eeqa
This gives  
equal effective masses for relative and CM motions. The  generalization for $N$ ions, see the Appendix A, 
also holds this property.   
The original coordinates and momenta are given by 
\beqa\label{2x2basis}
\begin{aligned}
\sf{q_1}&=&Q+r;\,\;{\sf{p_1}}=(P+p)/2, 
\\
\sf{q_2}&=&Q-r;\,\;{\sf{p_2}}=(P-p)/2.
\end{aligned}
\eeqa
Substituting Eq. (\ref{2x2basis}) in Eq. (\ref{H2har}), 
the Hamiltonian takes the form
\beqa
H(Q,P,r,p)&=&\frac{P^2}{2M}+\frac{1}{2}M\omega^2(Q-Q_0)^2
\nonumber\\
&+&\frac{p^2}{2M}+\frac{1}{2}M\omega^2r^2+\frac{C_c}{2r},
\eeqa
where $M=2m$ is the total mass. The Hamiltonian is the sum of two terms, $H=H_{cm}+H_r$, where each term depends only 
on one of the pairs coordinate-momentum. We may thus ``separate variables'' and find time-dependent solutions of the Schr\"odinger 
equation of the form 
\beqa
\Psi_{TOT}=\Psi_{cm}\otimes\Psi_r.
\eeqa
The relative part of the Hamiltonian,
\beq
H_r=\frac{p^2}{2M}+\frac{1}{2}M\omega^2r^2+\frac{C_c}{2r},
\eeq
does not depend on $Q_0(t)$ so the relative motion
is not affected by the transport and will remain unexcited.  
Thus we only need to design a trajectory for which the CM is unexcited at final time. 
This may be achieved adiabatically or via shortcuts-to-adiabaticity.  The CM Hamiltonian,
\beq
\label{CMH}
H_{cm}=\frac{P^2}{2M}+\frac{1}{2}M\omega^2[Q-Q_0(t)]^2,
\eeq
has the form of a particle of mass $M=2m$ in a harmonic trap, so any of the shortcut-to-adiabaticy
techniques known (using  Fast-Forward,  optimal control,  invariants, or their combination   \cite{Uli,fast-forward,Calarco,Erik})
may be applied to find a suitable $Q_0(t)$. 

To inverse engineer the trap trajectory making use of invariants, the invariant is designed first, 
consistent with a predetermined structure of the Hamiltonian  
\cite{Erik}. The invariant is parameterized by the 
classical trajectory $Q_c(t)$ that satisfies the
classical equation of motion $\ddot{Q}_c+\omega^2(Q_c-Q_0)=0$ 
and boundary conditions $Q_c(0)=\dot{Q}_c(0)=\ddot{Q}_c(0)=Q_0(0)=0$; 
$Q_c(t_f)=Q_0(t_f)=d$; $\dot{Q}_c(t_f)=\ddot{Q}_c(t_f)=0$. 
They  imply the initial and final commutativity between the invariant and the Hamiltonian, 
and the stability of the solution when the 
Hamiltonian remains constant beyond the boundary  times. 
A simple polynomial interpolation gives  
\cite{Erik}
\beqa
Q_c&=&d\left(10 s^3-15 s^4+6 s^5\right),
\nonumber\\
Q_0&=&\frac{d}{\omega^2t_f^2}\left(60 s-180 s^2+120 s^3\right)
\nonumber\\
&+&d\left(10 s^3-15 s^4+6 s^5\right),
\label{q0qc}
\eeqa
where $s=t/t_f$. 
Each initial eigenstate of $H_{cm}(0)$ would evolve exactly according to the ``transport mode''
\beq
\label{tm}
\Psi_n(Q,t)=e^{-\frac{i}{\hbar}[E_n t+\int_0^t \frac{M\dot{Q}_c^2}{2}dt']}e^{iM\dot{Q}_cQ/\hbar}\Phi_n(Q-Q_c),
\eeq
where
\beq   
\Phi_n(x)=\sqrt{\frac{1}{2^nn!}}\left(\frac{M\omega}{\pi\hbar}\right)^{1/4}e^{-\frac{M\omega x^2}{2\hbar}}H_n\left(\sqrt{\frac{M\omega}{\hbar}}x\right)
\eeq
are the eigenfunctions of a harmonic oscillator.
At $t_f$ the modes  become again eigenstates of the Hamiltonian $H(t_f)$
but at intermediate times they are in general a superposition of several eigenstates of $H(t)$. 
Note that, apart from transport between stationary states,  it is also possible to design  {launching} protocols, 
in which the system begins at rest and ends up with a 
given  center-of-mass velocity, and,   similarly,   stopping protocols \cite{Erik}.

The separability between CM and relative motions is still valid for two ions of different masses if they experience the same trapping 
frequency, but it breaks down  if the  frequency depends on position $Q_0$,  if the two ions experience different trapping frequencies, 
or in presence of anharmonicity. We shall concentrate on this later case, as it occurs in all traps and affects neutral atoms as well. 
\subsection{Anharmonic Trap}
We now consider an additional quartic potential in the Hamiltonian,
\beqa
H&=&\frac{{\sf p_1}^2}{2m}+\frac{{\sf p_2}^2}{2m}+\frac{C_c}{{\sf q_1}-{\sf q_2}}+\frac{1}{2}m\omega^2[({\sf q_1}-Q_0)^2
\nonumber\\
&+&({\sf q_2}-Q_0)^2+\beta({\sf q_1}-Q_0)^4+\beta({\sf q_2}-Q_0)^4],
\eeqa
where $\beta$ is a perturbative constant with dimensions $[L]^{-2}$ that sets  the ``strength''  of the  anharmonicity.
Non-rigid transport with a time-dependent trap frequency
or time-dependent anharmonicities due to noise or control limitations is clearly of interest but we shall only address here 
rigid transport as a first simpler step before considering more ambitious goals. 

In terms of CM and relative coordinates  we 
have
\beqa\label{hamcompleto}
H&=&\frac{P^2}{2M}+\frac{1}{2}M\omega^2[(Q-Q_0)^2+\beta(Q-Q_0)^4]
\nonumber\\
&+&\frac{p^2}{2M}+\frac{1}{2}M\omega^2(r^2+\beta r^4)+\frac{C_c}{2r}
\nonumber\\
&+&3M\omega^2\beta(Q-Q_0)^2 r^2
\nonumber\\
&=&H_{cm}+H_{r}+H_c.
\eeqa
The first two lines of Eq. (\ref{hamcompleto}) may be identified as (perturbed)  CM and relative Hamiltonians, $H_{cm}$ 
and $H_r$.    
Unlike the harmonic trap, there is now a coupling term $H_c$ (third line) that depends both on $Q$ and $r$
so the variables cannot be separated. No nontrivial invariants are known for this Hamiltonian \cite{quartic,invariants}, so
in principle we cannot inverse-engineer the trajectory exactly using invariants.
One approximate option is to design it for the unperturbed harmonic oscillator.   
An exact alternative is to apply a linear potential to compensate the inertial force as in \cite{Erik,Masuda,Sofia}.
\subsection{Compensating Force Approach}
In this subsection we introduce an additional time-dependent linear term in the  
Hamiltonian to compensate for the 
effect of the trap motion in the trap frame and avoid final excitations. 
This generalizes for two ions the results in \cite{Erik}. 
The extension of the compensating force approach to $N$ ions 
was discussed by Masuda in \cite{Masuda} using the Fast-Forward  approach, see also the Appendix. 
We shall use here (hat) operator notation since the commutativity of position and momentum plays a role. 

Let us 
first define a unitary transformation \cite{Erik, unitary} that shifts the momentum and position 
of the center of mass coordinate,  
\beq\label{desplacement}
\widehat{\mathcal U}=\widehat{\mathcal U}_1\widehat{\mathcal U}_2=e^{i\widehat{P}Q_0(t)/\hbar}e^{-iM\dot{Q}_0(t)\widehat{Q}/\hbar}.
\eeq
This amounts to change the reference system from a laboratory frame to the rest frame of the trap.\footnote{
Since $\widehat{P}$ and $\widehat{Q}$ do not commute  alternative orderings are possible but they only change the Hamiltonian 
by purely time-dependent terms without physical effect.}

We first rewrite the Hamiltonian in the lab frame as 
\beq
\widehat{H}(\widehat{Q}-Q_0,\widehat{r},\widehat{P},\widehat{p})=\frac{\widehat{P}^2}{2M}+\frac{\widehat{p}^2}{2M}+U(\widehat{Q}-Q_0,\widehat{r}). 
\eeq
The equation for the transformed (trap frame) wavefunction $|\Phi\rangle=\widehat{\mathcal{U}}|\Psi\rangle$, 
takes the form
\beqa
i\hbar\partial_t|\Phi\rangle&=&\widehat{\mathcal{U}}\widehat{H}\widehat{\mathcal{U}}^\dagger|\Phi\rangle
+i\hbar (\partial_t\widehat{\mathcal{U}})\widehat{\mathcal{U}}^ \dagger|\Phi\rangle
\nonumber\\
&=&\!\left[\!\widehat{H}(\widehat{Q},\widehat{r},\widehat{P},\widehat{p})\!+\!M(\widehat{Q}+Q_0)\ddot{Q}_0\!+\!\frac{1}{2}M\dot{Q}_0^2 \right]\!|\Phi\rangle\!.\nonumber\\
\eeqa
To compensate the inertial force $-M\ddot{Q}_0$ in the trap frame 
we may apply, in the lab-frame, a  force of the form  
\beq\label{force}
F=M\ddot{Q}_0,
\eeq
equivalently, a term $-M\widehat{Q}\ddot{Q}_0$ in the Hamiltonian. 
A further transformation $|\Phi'(t)\ra=\widehat{{\mathcal U}}'(t)|\Phi(t)\ra$, with 
$\widehat{{\mathcal U}}'(t)=e^{\frac{i}{\hbar}\int_0^t \frac{1}{2}M\dot{Q}_0^2 dt'}$ 
gives  
\beq\label{17}
i\hbar\partial_t|\Phi'\rangle=\left[\frac{\widehat{P}^2}{2M}+\frac{\widehat{p}^2}{2M}+\widehat{U}(\widehat{Q},\widehat{r})\right]|\Phi'\rangle.
\eeq
%
The resulting potential does not depend anymore on time, and any stationary state in the
rest frame of the trap will remain so during  transport. This holds for arbitrary 
potentials, even if $\widehat{Q}$ and $\widehat{r}$ 
are coupled, as in Eq. (\ref{hamcompleto}).  

A lower bound for the maximum acceleration of the compensating force is 
$2d/t_f^2$ \cite{Erik}. Since the forces that can be applied are   
typically limited by experimental
constraints, 
the compensation is not always easy to implement in practice, if at all.   
For this reason 
we study 
below alternative strategies. 
First we shall design the trap motion for an unperturbed harmonic potential
and analyze the effect of anharmonicity.  
\section{1D approximation\label{sec3}}
In this section we  discuss a simple 
approximation that provides valuable hints,  even in analytical form, on the transport behavior of two ions in 
presence of anharmonicities. The idea is to freeze the relative motion coordinate
at $r=r_e$, the minimum of the potential part that depends on $r$ only. 
Equivalently, we may consider a single-mode  approximation in which relative-motion excitations 
are neglected. Neglecting constant terms, the resulting Hamiltonian has the same form as the one for the frozen relative coordinate, 
substituting $r_e$ and $r_e^2$   
by the average values $\la r\ra$ and $\la r^2\ra$ in the ground relative-motion mode. With our parameters the relative differences  
are of order $10^{-3}$ which are not  
significant numerically so we use for simplicity the frozen values.    

With this assumption, and adding a constant term without physical effect,
the Hamiltonian (\ref{hamcompleto}) becomes 
\beq\label{h1d}
H=\frac{P^2}{2M}+\frac{1}{2}M\omega^2[(6\beta r_e^2+1)(Q-Q_0)^2+\beta(Q-Q_0)^4],  
\eeq
which we may also write as $H=H_0+\beta H_1$, where $H_1$ is a perturbation of the harmonic Hamiltonian $H_0$, 
\beq\label{h1}
H_1=\frac{1}{2}M\omega^2[6r_e^2(Q-Q_0)^2+(Q-Q_0)^4].
\eeq

\begin{figure}[t]
\begin{center}
\includegraphics[width=7.5cm]{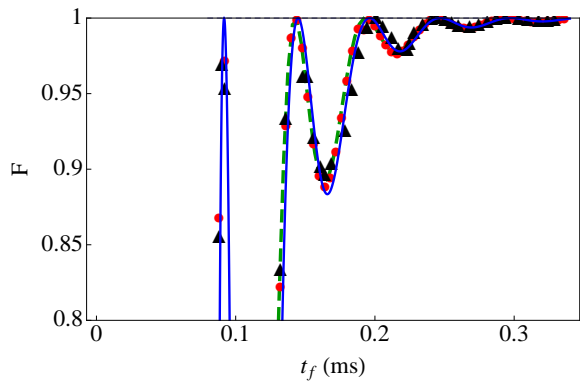}
\caption{\label{fig1}(Color online) Fidelity of the anharmonic system vs.  final time $t_f$ following the inverse engineering trajectory using 
second order perturbation theory (blue-thick line),  1D dynamics for the initial ground state of the harmonic oscillator (red-dotted line); 
1D dynamics for the initial ground state of the perturbed 1D Hamiltonian (green dashed line); 2D dynamics for the initial ground state of the 2D Hamiltonian (filled triangles). 
$M=2m=29.93\times 10^{-27}$ kg corresponding to ${}^{9}$Be${}^+$ ions, $\omega/(2\pi) =20$ kHz, $d=370$ $\mu$m, $r_e=62$ $\mu$m and $\beta=10^6$ m$^{-2}$.}
\end{center}
\end{figure}

Let the initial state 
be $|\Psi_n(0)\ra$. 
Using time-dependent perturbation theory the final wave vector $|{\Psi}(t_f)\ra$ is given by  
\cite{Erik,expansion,Andreas}
\beqa
&&|{\Psi}(t_f)\rangle=U_0(t_f,0)|{\Psi}_n(0)\rangle
\nonumber\\
&&{-\frac{i\beta}{\hbar}\int_{0}^{t_f}\!\! dt\,   U_0(t_f,t)H_1(t)|\Psi_n(t)\rangle}
\nonumber\\
&&-\frac{\beta^2}{\hbar^2}\int_0^{t_f}\!\! dt\! \int_0^t\!\! dt'\, U_0(t_f,t)H_1(t)U_0(t,t')H_1(t')|\Psi_n(t')\ra
\nonumber\\
&&+{\cal{O}}(\beta^3), 
\label{wf}
\eeqa
where $U_0$ is the unperturbed propagator for $H_0$. 
In terms of the complete set of transport modes, see Eq. (\ref{tm}),  it takes the form  
\beq
U_0(t,t')=\sum_j |\Psi_j(t)\ra\la\Psi_j(t')|.
\eeq 
%
To calculate the fidelity $F:=|\la \Psi_n(t_f)|\Psi(t_f)\ra|$ up to second order it is useful to separate the sum into $j=n$ and $j\ne n$ terms
in the second order contribution of Eq. (\ref{wf}).   
When computing $|\la\Psi_n(t_f)|\Psi(t_f)\ra|^2$ the square of first-order terms is cancelled by the second order term with $j=n$. 
Thus, the fidelity, up to second order,   
may finally be written as 
\beq 
\label{fidelity2}
F=\Bigg({1-\sum\limits_{j\neq n}|f_{j,n}^{(1)}|^2}\Bigg)^{1/2}, 
\eeq
where $f_{j,n}^{(1)}= \frac{-i\beta}{\hbar}\int_0^{t_f} dt\, \la \Psi_{j}(t)|H_1(t)|\Psi_n(t)\ra$. 
Due to the orthogonality properties of the Hermite polynomials, transitions induced by the quadratic perturbation will only be non-zero for one and two  level jumps. Instead, the quartic part of the perturbation will lead to jumps from one to four levels.  The 
$f_{j,n}^{(1)}$ transition amplitudes can be explicitly calculated 
so that the second order 
fidelity is  known analytically, although the form is too lengthy to be displayed here. Simplified expressions
will be provided later.   
We compare the fidelity in second order with the exact, numerical one 
(using the  Split-Operator method) in Fig. \ref{fig1}, starting both with the ground  state of the harmonic trap $\Phi_0(0)$ and the exact ground state 
of the anharmonic trap. The results are hardly distinguishable.
In the numerical examples we use the parameters in \cite{Bowler} except for a lower trap frequency to enhance anharmonic effects. 
The trap trajectory $Q_0(t)$ is chosen as in Eq. (\ref{q0qc}), using invariant-based engineering for the unperturbed system
with a polynomial ansatz for $Q_c$.   
The fidelity oscillates, 
reaching the maximum value of one at discrete values of  $t_f$. 
The occurrence of maxima is a generic feature that does not depend on the specific value of $\beta$ chosen. 
Below we shall work out a theory to explain and predict them.

%
%
%

%
\begin{figure}[t]
\begin{center}
\includegraphics[width=7.5cm]{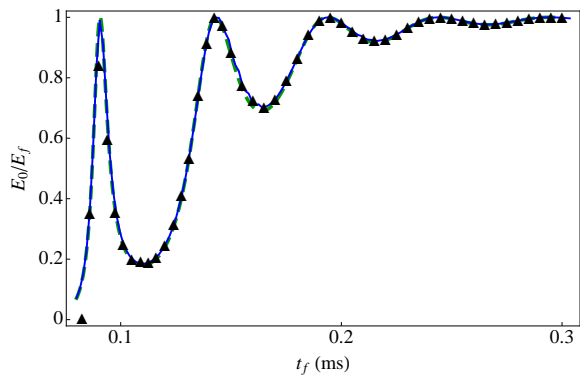}
\caption{\label{fig5}(Color online) 
For 1D calculations (blue solid line) we 
plot the ratio of initial, $E_0$ -the ground state energy-,  and final energies $E_f$, using  $H$ in Eq. (\ref{h1d}).    
$E_0/(E_{ex}(t_f)+E_0)$  (green-dashed line, hardly distinguishable from the previous line)
 is also plotted from a single classical trajectory, see Eq. (\ref{energiaexc}).
For 2D  (filled triangles)  we use Eq. (\ref{hamcompleto}) and redefine $E_0=\la H(0)\ra-\delta$, and $E_f=\la H(t_f)\ra-\delta$
where $\delta\equiv E_r^{(0)}+{\rm{min}}[V(0)]$. $E_r^{(0)}$ is the ground state of $H_r$, see Eq. (\ref{hamcompleto}), 
and $\rm{min}[V(0)]$
is the minimum of the potential  part in Eq. (\ref{hamcompleto}). The subtraction of $\delta$ provides results 
directly comparable with the 1D calculations. 
Same parameters as in Fig. 1.} 
\end{center}
\end{figure}
%

We shall now study the effect of each perturbation separately. The quadratic perturbation 
amounts to having designed the trap trajectory with the ``wrong'' trap frequency and, as we shall see, 
is the dominant perturbation except for very short times. The influence of the quartic perturbation was 
analyzed in \cite{Erik} but only with a much less accurate first-order approach. 
%
%
\begin{figure}[t]
\begin{center}
\includegraphics[width=4.2cm]{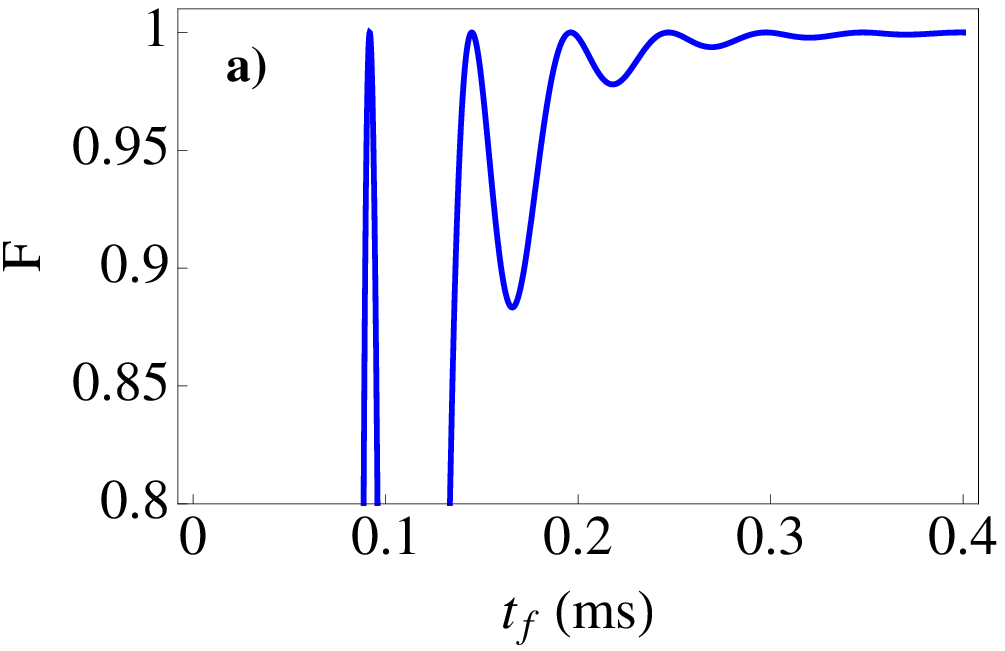}
\includegraphics[width=4.2cm]{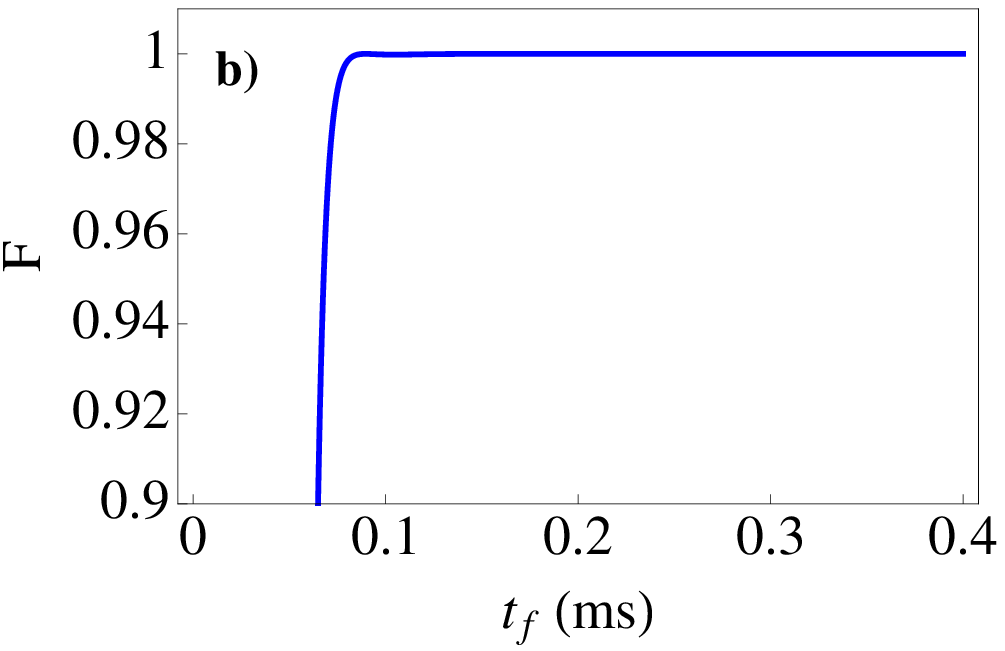}
\caption{\label{fig3}(Color online) Fidelity vs. final time ($t_f$) for the second order perturbation theory, 
indistinguishable from an exact 
1D quantum dynamical calculation (The initial state is the ground state of the perturbed harmonic oscillator) 
for a) the  quadratic perturbation in Eq. (\ref{h1}),  and b) the quartic perturbation in Eq. (\ref{h1}).
Same parameters as in Fig. 1.}
\end{center}
\end{figure}
%
The effect of the two perturbations is quite different as seen in Fig. \ref{fig3}.   
The quadratic perturbation  provides a fidelity almost identical to that of the total perturbation,
reproducing its oscillations and {peak times}. The quartic perturbation alone leads to a sudden growth in the fidelity
around a critical time $t_f^{cr}$,  
followed by fidelity $1$ for longer final times. To estimate the behavior of $t_f^{cr}$  with respect to  
transport and potential parameters
we note that the maximum of $|Q_{c}(t)-Q_0(t)|$
is  ${10 d}/{(\omega^2 t_f^2 3^{1/2})}$. Comparing the quadratic and quartic  contributions to the potential there 
we get 
\beq\label{tcri}
t_f^{cr}=\alpha\, \frac{\beta^{1/4} d^{1/2}}{\omega},
\eeq
where $\alpha\approx 16.5$ is adjusted numerically. 
For the parameters of Fig. \ref{fig3} this transition is to the right of the first peak of the quadratic perturbation
so the effect of the quartic perturbation is negligible.    
 
Let us now analyze in more detail the quadratic perturbation alone. It implies one and two vibrational quanta
as mentioned before.  
If we consider only  $n\rightarrow n\pm1$ the results are already very similar to the fidelity in Fig. \ref{fig3} (a). 
Since one-level transitions are dominant we can write down an explicit approximate 
form for the fidelity based on them,
\beqa
&&f_{n\pm1,n}^{(1)}=\frac{\pm360 i d\beta r_e^2 e^{\mp\frac{1}{2}it_f\omega}\sqrt{2(1+n)M\hbar}}{t_f^5\omega^{9/2}}
\nonumber\\
&\times&\left[6t_f\omega\cos\left(\frac{t_f\omega}{2}\right)+(t_f^2\omega^2-12)\sin\left(\frac{t_f\omega}{2}\right)\right]\!\!,
\label{fnn}
\eeqa
note the square root scaling with the mass. 
This amplitude is zero,  and the fidelity one,  when 
\beq\label{condicion1}
6t_f\omega\cos\left(\frac{t_f\omega}{2}\right)+(t_f^2\omega^2-12)\sin\left(\frac{t_f\omega}{2}\right)=0. 
\eeq
There is a $\beta$-independent solution for, approximately, every oscillation period. 
This result also follows from a simple classical argument:  
Consider a classical trajectory $\widetilde{Q}_c(t)$ satisfying 
\beq
\frac{\ddot{\widetilde{Q}}_c}{\widetilde{\omega}^2}+\widetilde{Q}_c-Q_0(\omega)=0,
\eeq
where $Q_0(\omega)=Q_0(t;\omega)$ is the trap trajectory calculated as before with $\omega$, Eq. (\ref{q0qc}), and  
$\tilde{\omega}=\omega\sqrt{1+6\beta r_e^2}$ is an effective trap frequency, shifted 
with respect to $\omega$ because of the relative-CM coupling, see Eq. (\ref{h1d}). 
Its energy for $\widetilde{Q}_c(0)=\dot{\widetilde{Q}}_c(0)=0$ is given by 
\beq\label{energiaexc}
E_{ex}(t)=\frac{1}{2}M\dot{\widetilde{Q}}_c^2(t)+\frac{1}{2}M\widetilde{\omega}^2\left[\widetilde{Q}_c(t)-Q_0(t)\right]^2.
\eeq
At time $t_f$ 
we have  
%
\beqa\label{excitation}
&&E_{ex}(t_f)=\frac{7200d^2M(\omega^2-\tilde{\omega}^2)^2}{t_f^{10}\omega^4\tilde{\omega}^8}
\nonumber\\
&\times&\!\!\left[6t_f\tilde{\omega}\cos\!\left(\frac{t_f\tilde{\omega}}{2}\right)\!+\!(t_f^2\tilde{\omega}^2-12)\sin\!\left(\frac{t_f\tilde{\omega}}{2}\right)\right]^2\!\!.
\eeqa
The condition for a zero is the same as Eq. (\ref{condicion1}) substituting $\omega\to\tilde{\omega}$. 
This leads to a very small displacement (and dependence on $\beta$) of the zeros for our parameters.
In Fig. \ref{fig5} we represent $E_0/(E_0+E_{ex}(t_f))$ which is indistinguishable from the  quantum 
curve for $E_0/E_f$.    
We may conclude unambiguously that the oscillations are  not quantum in nature.

\begin{figure}[h]
\begin{center}
\includegraphics[width=7.5cm]{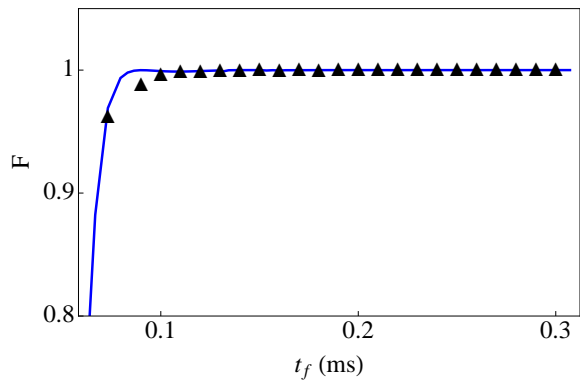}
\caption{\label{fig6}(Color online)
Fidelity vs. final time $t_f$ for adjusted trap trajectories $Q_0(t;\tilde{\omega})$.  
The initial condition is the ground state. 
1D:  blue solid line;  2D:  filled triangles.
Same parameters as in Fig. 1.} 
\end{center}
\end{figure}

Rather than adjusting the transport time to the discrete set of zeros,  a better, more robust strategy that allows for  a
continuous set of final  times is to design the trap trajectory 
taking into account the frequency shift. Changing $\omega\to\tilde{\omega}$ in Eq. (\ref{q0qc})
we get an adjusted trajectory $Q_0(t;\tilde{\omega})$ for which  
$E_{ex}(t_f)=0$ by construction for any $t_f$. 
Similarly, $Q_0(t;\tilde{\omega})$
gives  fidelity one for all $t_f$ in the 1D model,  if only the quadratic perturbation is considered. 
In the protocol based on $Q_0(t;\tilde{\omega})$ the only disturbance comes from the quartic term that sets the speed limitation given by 
Eq. (\ref{tcri}).  Fig. \ref{fig6} shows the impressive results of this simple approach. 
In practice $\tilde{\omega}/(2\pi)$ may be measured as the effective CM-mode frequency.
%

\begin{figure}[h]
\begin{center}
\includegraphics[width=7.5cm]{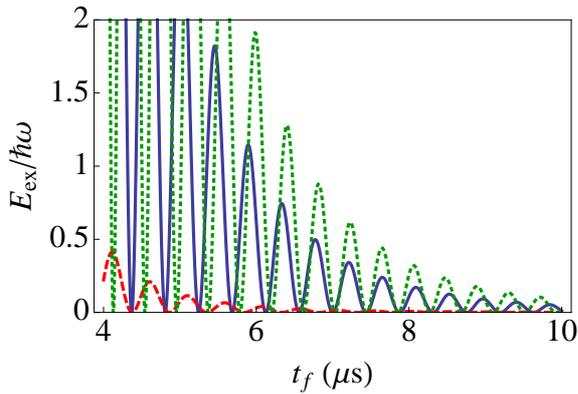}
\caption{\label{new}(Color online) Motional excitation quanta vs. final time. 
$M=2m=29.93 \times 10^{-27}$ kg, ${\omega}/({2\pi})=2$ MHz, 
$d=370$ $\mu$m for the three cases and $\beta=6.4\times 10^9$ m$^{-2}$, $r_e=2.807 \mu$m (solid blue line), $\beta=10^9$ m$^{-2}$, $r_e=2.883$ $\mu$m (red-dashed line) and $\beta=10^{10}$ m$^{-2}$, $r_e=2.764$ $\mu$m (green-dotted line).
The trap trajectory is given by Eq. (\ref{q0qc}). If instead the adjusted trajectory $Q_0(t;\tilde{\omega})$
is used, then $E_{ex}(t_f)=0.$} 
\end{center}
\end{figure}

Higher, more realistic  trap frequencies lead to similar results but for a larger $\beta$.  
Simple estimates of the fidelity or excitation may be drawn from Eqs. (\ref{fnn}) or (\ref{energiaexc}). Figure \ref{new} depicts the 
classical excitation energy 
of Eq. (\ref{energiaexc}) for $\omega/(2\pi)=2$ MHz and different values of $\beta$ using the (unshifted) $\omega$ 
in $Q_0(t)$. Notice that for these large $\beta$ values the times of minimum excitation do change with $\beta$. 
The middle value of $\beta$ is chosen so that at $t_f=8$ $\mu$s the excitation is similar to the one seen experimentally 
in \cite{Bowler}. For the adjusted trajectory $Q_0(t;\tilde{\omega})$,  $E_{ex}(t_f)=0$ as before. 
%
%
%
%
%
%
%
%
\section{Full 2D analysis\label{sec4}}
We have also examined the evolution of the state according to the full 2-dimensional Hamiltonian (\ref{hamcompleto}), without freezing  the relative 
motion, using a 2D split-operator method to simulate  quantum 
dynamics. 
The computation is performed in the trap frame to reduce the numerical grid size. 
Figure \ref{fig1} shows that the quantum fidelities of the 1D model are in very good agreement with the fidelities  calculated for 2D dynamics. 
Figure \ref{fig5} shows energy  ratios for 1D and 2D calculations. To compare them on equal footing 
in 2D the minimum of the potential and the ground state relative energy are subtracted, see the caption for details.
Again the 1D and 2D quantum calculations are remarkably close to each other. 
2D calculations may also be found in Fig. \ref{fig6} for the transport designed using a shifted frequency. They confirm the 
excellent performance of this strategy with respect to the anharmonic perturbation.      
\section{Discussion\label{sec6}}
For two ions in a harmonic trap the relative motion is uncoupled to the CM motion. 
They may be transported faster than adiabatically treating the center of mass (CM) as a single particle 
and applying different shortcuts to adiabaticity.  
For anharmonic traps CM and relative motion are coupled. 
A 1D model for the CM has been first worked out based on a single relative-motion mode, or, equivalently, 
freezing the relative coordinate. The full 2D quantum calculations show excellent agreement with 
this model in the parameter range studied.  
It is possible to achieve fast and faithful transport for an arbitrary trap shape by compensating for the
inertial force in the trap frame with a linear potential. 
That may be difficult in practice so other strategies to get  high fidelities have been explored.   
For a quartic anharmonicity the effective 1D potential includes a quartic and a quadratic perturbation.
The latter is usually dominant except for very short transport times. 
If the trap trajectory is the one designed for the unperturbed (harmonic) trap, the quartic perturbation 
alone implies a  sharp increase
to one of the fidelity, while  the quadratic perturbation induces (classical) fidelity oscillations
with respect to the final time $t_f$.
Taking into account the shift in the effective trap frequency due to the coupling, the 
trap trajectory is much more robust and the effect of the quadratic perturbation is cancelled. 
 
Other aspects worth investigating for future work are the effects of different types of noise \cite{Andreas}, 
and other anharmonic forms such as  cubic or time-dependent perturbations \cite{Uli,anh}.  
Variations of the trap frequency $\omega$ and anharmonicity factor $\beta$ with time could be affected
by random and/or systematic perturbations. We 
expect that techniques similar to the ones applied in \cite{Andreas,dephasing} to design robust trap trajectories
will be instrumental in designing robot trajectories.  
We also intend to other relevant systems such as pairs of different ions as well as transport of  four or more ions \cite{4}.
For more than three ions quantum computations are quite challenging but classical methods should provide a 
good guidance, as shown here for two ions.    
\section*{Acknowledgements}
We are grateful to  D. Leibfried and U. Poschinger for useful comments. 
We acknowledge funding by Grants No. IT472-10, 
FIS2009-12773-C02-01, and  
the UPV/EHU Program
UFI 11/55. M. P. acknowledges a fellowship by UPV/EHU.
\section*{Appendix A: N-Ion transport\label{sec5}}
We shall first address the transport of $N$ equal ions in a single harmonic trap. 
The compensating force approach for an arbitrary trap of Sec. III C  will be 
generalized afterwards. 

In a harmonic trap the Hamiltonian is given by $N$ coordinates for the positions of each of the ions (${\sf q_1}$, ${\sf q_2}$, ${\sf q_3}$, $\dots$, $\sf{q_N}$), and the corresponding momenta, 
\beqa
H(\{{\sf q_i}, {\sf p_i}\})&=&\frac{1}{2m}\sum\limits_{i=1}^{N}{\sf {p}_i}^2+\frac{1}{2}m\omega^2\sum\limits_{i=1}^{N}({\sf q_i}-Q_0)^2
\nonumber\\
&+&{\sum_{i=1}^{N-1}\sum_{j=i+1}^{N}}\frac{C_c}{{\sf q_i}-{\sf q_j}},
\nonumber
\eeqa
where ${\sf q_1}>{\sf q_2}>\cdots>{\sf q_{N-1}}>{\sf q_N}$ because of  the strong Coulomb repulsion.
We now define a CM and relative coordinates and momenta,
\beqa
\nonumber
&&\!\!\!\!\!\!\!\!\!\!Q=\frac{1}{N}\sum_{i=1}^N {\sf q_i},\;\;  P=\sum_{i=1}^{N} {\sf p_i},
\\
&&\!\!\!\!\!\!\!\!\!\!r_{i}=\frac{\sf{q_i}-{\sf{q_{i+1}}}}{N},\;  p_{i}={\sf p_i}-{\sf{p_{i+1}}},\; i=1,2,\dots, N-1,
\nonumber
\eeqa
corresponding to the inverse transformation 
\beqa
{\sf q_i}&=&Q+\sum_{j=1}^{N-i} jr_{N-j} -\sum_{k=1}^{i-1} kr_{k},
\nonumber\\
{\sf p_i}&=&P+\frac{1}{N}\sum_{j=1}^{N-i} j p_{{N-j}} -\frac{1}{N}\sum_{k=1}^{i-1} kp_{k}.
\nonumber
\eeqa

The Hamiltonian in  the new coordinates is
\beqa
H&=&\frac{P^2}{2M}+\frac{1}{2}M\omega^2(Q-Q_0)^2
\nonumber\\
&+&\frac{1}{2NM}\sum_{i=1}^{N}\Bigg[\Bigg(\sum_{j=1}^{N-i} j p_{N-j} \Bigg)^{\!\!2} + \left(\sum_{k=1}^{i-1}k p_{k}\right)^{\!\!2}
\nonumber\\
&-&2\sum_{j=1}^{N-i} \sum_{k=1}^{i-1} j k p_{N-j} p_{k} \Bigg]
\nonumber\\
&+&\frac{1}{2N}M\omega^2\sum_{i=1}^{N}\Bigg[\Bigg(\sum_{j=1}^{N-i}jr_{N-j}\Bigg)^{\!\!2}+\left(\sum_{k=1}^{i-1}kr_{k}\right)^{\!\!2}
\nonumber\\
&-&2\sum_{j=1}^{N-i}\sum_{k=1}^{i-1}jkr_{N-j}r_{k}\Bigg]
\nonumber\\
&+&\frac{C_c}{N}\left(\sum_{i=1}^{N-1}\frac{1}{r_i}+\sum_{i=1}^{N-2}\sum_{j=i+1}^{N-1} \frac{1}{\sum_{k=i}^j r_k}\right),
\nonumber
\eeqa
where $M=Nm$. 
As for two ions, the Hamiltonian can be written as the sum of two terms, 
$$
H=H_{cm}(Q,P)+H_r(\{r_i,p_i\}),
$$
where $H_{cm}$ has the same form as Eq. (\ref{CMH}), 
and $H_r$ depends only on $N-1$ relative coordinates and their corresponding momenta. 
It does not depend on the trap 
trajectory $Q_0(t)$, so this system can be  transported without final excitations by following any shortcut-to-adiabaticity trap trajectory 
for a particle of mass $M$. 

For an arbitrary potential $U(Q-Q_0,\{r_i\})$ a compensating force like the one in Eq. (\ref{force}), i.e.,  
\beq
\nonumber
H=\frac{P^2}{2M}+K(\{p_i\})+U(Q,\{r_{i}\})-MQ\ddot{Q}_0,
\eeq
where $K(\{p_i\})$ is the relative kinetic energy, 
provides in the trap frame, using  Eq. (\ref{desplacement}),
\beq
\nonumber
i\hbar\partial_t|\Phi\rangle=\left[\frac{P^2}{2M}+K(\{p_i\})+U(Q,\{r_i\})+\frac{M\dot{Q}_0^2}{2}\right]|\Phi\rangle.
\eeq
The time-dependent term is not physically significant and may be eliminated as in Eq. (\ref{17}). 
As for two ions, any stationary state will remain unperturbed thanks to the compensating force. 
For a different derivation of this result see 
\cite{Masuda}.
%
%
%
%
%

%
%
%
%
%
%
%
%
%
%
%
%
%
%
%
%
%

%
%

\begin{thebibliography}{10}
%
\bibitem{q.computer} D. Kielpinski, C. Monroe and D. J. Wineland Nature \textbf{417}, 709 (2002).
%
\bibitem{Rowe} M. A. Rowe et al., Quantum Inf. Comput. \textbf{4}, 257 (2002).
%
\bibitem{Wineland} R. Reichle, D. Leibfried, R. B. Blakestad, J. Britton, J. D. Jost, E. Knill, C. Langer, R. Ozeri, S. Seidelin, and D. J. Wineland, Fortschr. Phys. \textbf{54}, 666 (2006).
%
\bibitem{simulator} H.-K. Lau and D. F. V. James, Phys. Rev. A \textbf{85}, 062329 (2012).
%
\bibitem{Uli} S. Schulz, U. Poschinger, K. Singer, and F. Schmidt-Kaler, 
Fortschr. Phys. \textbf{54}, 648 (2006).
%
\bibitem{BECtransport} E. Torrontegui, X. Chen, M. Modugno, S. Schmidt, A. Ruschhaupt, and J. G. Muga, New J. Phys. \textbf{14}, 013031 (2012).
%
\bibitem{David} A. Couvert, T. Kawalec, G. Reinaudi, and D. Gu\'ery-Odelin, Europhys. Lett. \textbf{83}, 13001 (2008).
%
\bibitem{Calarco} M. Murphy, L. Jiang, N. Khaneja, and T. Calarco, Phys. Rev. A \textbf{79}, 020301(R) (2009).
%
\bibitem{fast-forward} S. Masuda and K. Nakamura, Proc. R. Soc. A \textbf{466}, 1135 (2010).
%
\bibitem{Erik} E. Torrontegui, S. Ib\'a\~nez, X. Chen, A. Ruschhaupt, D. Gu\'ery-Odelin, and J. G. Muga, Phys. Rev. A \textbf{83}, 013415 (2011).
%
\bibitem{Xi} X. Chen, E. Torrontegui, D. Stefanatos, Jr-Shin Li, and J. G. Muga, Phys. Rev. A \textbf{84}, 043415 (2011).
%
\bibitem{Bowler} R. Bowler, J. Gaebler, Y. Lin, T.  R. Tan, D. Hanneke, J. D. Jost, J.  P. Home, D. Leibfried, and D.  J. Wineland, Phys. Rev. Lett. \textbf{109}, 080502 (2012). 
%
\bibitem{Schmidt-Kaler} A. Walther, F. Ziesel, T. Ruster, S. T. Dawkins, K. Ott, M. Hettrich, K. Singer, F. Schmidt-Kaler, and U. Poschinger, Phys. Rev. Lett. \textbf{109}, 080501 (2012).
%
\bibitem{Roos} C. Roos, Physics \textbf{5}, 94 (2012).
\bibitem{review} E. Torrontegui, S. Ib\'a\~nez, S. Mart\'\i nez-Garaot,
M. Modugno, A.  del Campo, D. Gu\'ery-Odelin, A. Ruschhaupt, X. Chen,
and J. G.  Muga, Adv. At. Mol. Opt. Phys. \textbf{62}, 117 (2013).
%
\bibitem{Andreas} A. Ruschhaupt, X. Chen, D. Alonso, and J. G. Muga, New J. Phys. \textbf{14}, 093040 (2012).
%
\bibitem{anh} J. P. Home, D. Hanneke, J. D. Jost, D. Leibfried, and D. J. Wineland, 
New J. Phys. \textbf{13}, 073026 (2011).
\bibitem{quartic} N. W. Evans, J.  Math. Phys. \textbf{31}, 600 (1990).
%
\bibitem{invariants} M. Karlovini, G. Pucacco, K. Rosquist, and L. Samuelsson, J. Math. Phys. \textbf{43}, 4041 (2002).
%
\bibitem{Masuda} S. Masuda, Phys.Rev. A \textbf{86}, 063624 (2012).
%
\bibitem{Sofia} E. Torrontegui, S. Mart\'\i nez-Garaot, A. Ruschhaupt, and J. G. Muga, Phys. Rev. A \textbf{86}, 013601 (2012).
%
\bibitem{unitary} W. H. Klink, Proceedings of the Second International Conference on Symmetry in Nonlinear Mathematical Physics {\textbf{2}}, 254 (1997).
%
\bibitem{expansion} E. Torrontegui, X. Chen, M. Modugno, A. Ruschhaupt, D. Gu\'ery-Odelin, and J. G. Muga, Phys. Rev. A \textbf{85}, 033605 (2012).

%

\bibitem{dephasing} X.-J. Lu, X. Chen, A. Ruschhaupt, D. Alonso, S. Gu\'erin, and J. G. Muga, 
Phys. Rev. A \textbf{88}, 033406 (2013).
%
\bibitem{4}J.  P. Home, D. Hanneke, J. D. Jost, J. M. Amini,
D. Leibfried, and D. J. Wineland, Science \textbf{325}, 1227 (2009). 
%



\end{thebibliography}
\end{document}